\documentclass[%
 reprint,amsmath,amssymb,aps]{revtex4-2}
 \usepackage{graphicx}% Include figure files
\usepackage{dcolumn}% Align table columns on decimal point
\usepackage[export]{adjustbox}
\usepackage{bm}% bold math
\usepackage{tabularx,multirow,array,booktabs,makecell,float}
\usepackage[utf8]{inputenc}
\usepackage[T1]{fontenc}

\usepackage[dvipsnames]{xcolor} % colours!!!!
\usepackage[normalem]{ulem} % for striking through text horizontally
\begin{document}

\title{
High-dimensional entanglement for quantum communication in the frequency domain}

% Author: Please give full first and last names for authors and include * after the name of all corresponding authors

\author{Meritxell Cabrejo-Ponce}
\email{meritxell.cabrejo.ponce@iof.fraunhofer.de}
\affiliation{Fraunhofer Institute for Applied Optics and Precision Engineering, Albert-Einstein-Strasse 7, 07745 Jena}
\affiliation{Abbe Center of PhotonicsFriedrich Schiller University JenaAlbert-Einstein-Str. 6, 07745 Jena, Germany}%

\author{André Luiz Marques Muniz}
\affiliation{Fraunhofer Institute for Applied Optics and Precision Engineering, Albert-Einstein-Strasse 7, 07745 Jena
}%

\author{Marcus Huber}
\email{marcus.huber@tuwien.ac.at}
\affiliation{Atominstitut,  Technische  Universit{\"a}t  Wien,  Stadionallee 2, 1020  Vienna,  Austria}
\affiliation{Institute for Quantum Optics and Quantum Information (IQOQI), Austrian Academy of Sciences, 1090 Vienna, Austria
}
\author{Fabian Steinlechner}
\email{fabian.steinlechner@iof.fraunhofer.de}
\affiliation{Fraunhofer Institute for Applied Optics and Precision Engineering, Albert-Einstein-Strasse 7, 07745 Jena}
\affiliation{Abbe Center of Photonics, Friedrich Schiller University Jena, Albert-Einstein-Str. 6, 07745 Jena, Germany
}%

% Keywords: Please provide a minimum of three and a maximum of seven keywords, separated by commas
%\keywords{quantum frequency combs, entanglement, high-dimensional, quantum information}

% Abstract should be written in the present tense and impersonal style (i.e., avoid we), and be at most 200 words long
\begin{abstract}
High-dimensional photonic entanglement is a promising candidate for error-protected quantum information processing with improved capacity. Encoding high-dimensional qudits in the carrier frequency of photons combines ease of generation, universal single-photon gates, and compatibility with fiber transmission for high-capacity quantum communication. Recent landmark experiments have impressively demonstrated quantum interference of a few frequency modes, yet the certification of massive-dimensional frequency entanglement has remained an open challenge. This study shows how to harness the large frequency-entanglement inherent in standard continuous-wave spontaneous parametric down-conversion processes. It further reports a record certification of discretized frequency entanglement, combined with a novel approach for certification that is both highly efficient and nonlocally implementable. This technique requires very few measurements and does not require assumptions on the state. The work opens the possibility for utilizing this encoding in quantum communications and in quantum information science in general.
\end{abstract}
\maketitle

% Text: Please use section headings and subheadings as specified below. For communications, all section headings apart from Experimental Section should be removed
% Please make the first reference to a display item bold: \textbf{Figure 1}
% Do not abbreviate Figure, Equation, etc.; display items are always singular, i.e., Figure 1 and 2.
% Equations are always singular, i.e., Equation 1 and 2, and should be inserted using the {equation} environment, not as graphics
% Please do not use footnotes in the text, additional information can be added to the Reference list.

\section{Introduction}
Entanglement is a unique and powerful quantum feature with a multitude of applications in quantum information processing. The nonlocal correlations of the entangled states may be used in quantum communications, imaging, metrology, and quantum processors. 
In the case of photons, polarization-entangled states have been traditionally used to demonstrate a multitude of quantum gates and quantum information protocols \cite{kokLinearOpticalQuantum2007} and the principles of rapidly developing quantum networks \cite{panMultiphotonEntanglementInterferometry2012,anwarEntangledPhotonpairSources2021}. These qubit states are easy to manipulate with linear optics and can be distributed in fiber or free-space links because of their low interaction with the environment.
Nevertheless, larger alphabets in quantum communication are highly pursued not only to increase the capacity of the quantum channel.
High-dimensional encoding also provides a stronger tolerance to noise \cite{eckerOvercomingNoiseEntanglement2019}, an essential asset to overcome the transmission limits of polarization qubits. On the other hand, an increase in dimensionality can also boost the computational power of quantum computers \cite{lanyonSimplifyingQuantumLogic2009}. In this context, a recent landmark experiment \cite{vigliarErrorprotectedQubitsSilicon2021} used high-dimensional entanglement of propagation paths on a silicon photonic chip to realise error-protected logical qubits and thus improve the performance of a quantum phase estimation algorithm. Harnessing the full state space of photonic degrees of freedom (DOF) such as transverse spatial mode, time, and frequency, will be key to future generations of photonic quantum information processors.

\begin{figure*}[t!]
\centering
\includegraphics[width=1\textwidth]{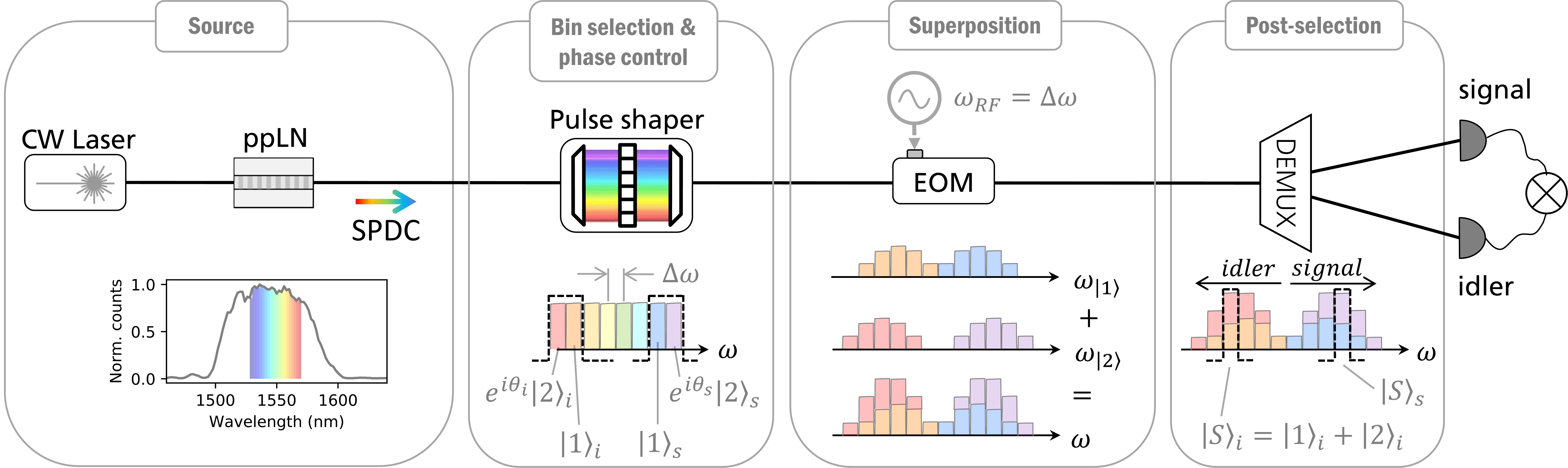}
\caption{Schematic of our frequency-entangled photon pair source and the state analysis parts. We generate broadband frequency entanglement at telecom wavelength via spontaneous parametric down-conversion (SPDC) in a periodically poled lithium niobate waveguide (ppLN). Afterwards, the photon pairs are coherently manipulated with a pulse shaper and an electro-optic modulator (EOM) to analyze the correlations in a superposition basis. CW: continuous-wave; DEMUX: demultiplexer.}
\label{fig:Setup}
\end{figure*}

Although spatial \cite{mairEntanglementOrbitalAngular2001,krennGenerationConfirmation1002014,fontaineLaguerreGaussianModeSorter2019} and temporal modes \cite{fransonBellInequalityPosition1989,brendelPulsedEnergyTimeEntangled1999,richartExperimentalImplementationHigher2012} have been straightforward to operate in very large dimensional spaces, the frequency DOF has remained behind in these advances. The frequency domain at optical scales is of particular interest because of its parallelization capabilities, in particular its compatibility with telecom multiplexing or frequency modulation techniques. Earlier experiments exploiting electro-optic modulation were able to demonstrate two-dimensional frequency entanglement \cite{olislagerImplementingTwophotonInterference2012}, while the use of pulse shapers and up-conversion processes allowed the characterization of up to four-dimensional states \cite{peerTemporalShapingEntangled2005,bernhardShapingFrequencyentangledQudits2013}. 
The combination of both components was used later to coherently control the quantum states emerging from integrated resonators \cite{kuesOnchipGenerationHighdimensional2017}.
These important building blocks opened the possibility to measure in the superposition basis at optical scales and were used to demonstrate discretized frequency entanglement with few dimensions (up to 6 \cite{luBayesianTomographyHighdimensional2022}). 
Since then, approaches have been developed to perform arbitrary manipulations of single-frequency qubits \cite{luFullyArbitraryControl2020} and the first steps have been taken towards full control of single qudits, its high-dimensional version 
\cite{luQuantumInterferenceCorrelation2018,luControlledNOTGateFrequencybin2019,cuiHighDimensionalFrequencyEncodedQuantum2020}. 
While Hong-Ou-Mandel interference has also been used to verify two-dimensional frequency entanglement \cite{chenVerificationHighdimensionalEntanglement2020}, the requirement of local measurements limits the utility of these methods in quantum networks.

Furthermore, it was recently shown that the HOM based approach cannot be used to verify higher dimensional entanglement \cite{lingarajuQuantumFrequencyCombs2019}
, at least not without imposing additional assumptions such as energy conservation \cite{chenTemporalDistinguishabilityHongOuMandel2021}.

Entangled frequency states can also be generated in combination with entanglement in other DOFs like polarization, path \cite{zhangliangOnchipParallelProcessing2021} or even temporal, whenever the relevant time and frequency properties can be manipulated independently, because they refer to vastly different time scales. 
Such hyperentangled states can also be used to enlarge the dimensionality of the system, to generate cluster states \cite{reimerHighdimensionalOnewayQuantum2019} or to perform more advanced quantum gates \cite{imanyHighdimensionalOpticalQuantum2019}.
Yet, all these approaches are only able to manipulate a small frequency mode set of typically extensive underlying quantum states. 
Certifying real high-dimensional entanglement is not trivial, particularly in the frequency domain, and its immense potential is still unexploited.

In this Letter, we show that it is not always necessary to design quantum sources with a discretized frequency space, such as those built in cavities, and therefore continuous spectra can also provide access to massive-dimensional and well-controlled Hilbert space. We show quantum interference with up to 7 modes and $>98\%$ visibility and demonstrate complete state control in a 7-mode state space. Subsequently, based on previous work to characterize time-bin qudits \cite{martinQuantifyingPhotonicHighDimensional2017,tiranovQuantificationMultidimensionalEntanglement2017}, we certify genuine high-dimensional frequency entanglement without prior assumptions regarding, e.g., the purity of the quantum state.
We can further alleviate the harsh requirements on the number of measurements necessary to characterize the relevant correlations, which grows with dimensionality, by taking into account energy-conservation. This step is not necessary for our certification method, but it allows us exploit a novel bucket detection approach that requires very few measurement settings, in a similar fashion to the compressed measurements used to characterize spatial correlations \cite{schneelochQuantifyingHighdimensionalEntanglement2018,schneelochQuantifyingEntanglement68billiondimensional2019}. 
Finally, by recovering the information from high-quality quantum interference of two-dimensional (2D) subspaces, we are able to certify a minimum of 33 entangled frequency modes, the highest dimensionality of entanglement reported in time and frequency degrees of freedom.

\section{Results and Discussion}
\subsection{Frequency comb generation}
In our work, we analyze the frequency content of two-photon states generated in standard $\chi^{(2)}$ non-linear crystals: periodically poled lithium niobate (ppLN) waveguides (see Experimental Section). 
They provide continuous and broadband spontaneous parametric down-converted photons (SPDC) with high efficiency and 60 nm of bandwidth. 
The SPDC process is temperature tuned to a degeneracy wavelength of 1548 nm to cover the entire C band with high uniformity (see inset in Fig. \ref{fig:Setup}). 

At this point, the frequency space of the generated photon pairs would typically be discretized, e.g., with etalon cavities to carve the spectrum \cite{xieHarnessingHighdimensionalHyperentanglement2015,imanyCharacterizationCoherentQuantum2018}. 
This approach is useful for the isolation of frequency modes and for performing sideband modulation. 
However, a considerable contribution of the photon spectrum is directly rejected, reducing the total throughput. 
For this reason, we avoid this step and employ the maximum bandwidth per frequency mode. 

Due to the energy conservation of the SPDC process, the emitted photon pairs are strongly anticorrelated in frequency, and the state can be described as:

\begin{equation}
    |j\rangle_s |j\rangle_i = \int \Pi(\Omega -j\Delta\omega, \Omega +j\Delta\omega)|\omega_0 + \Omega\rangle_s |\omega_0 - \Omega\rangle_i \,d\Omega
\end{equation}
where $|j\rangle_{s,i}$ is the label of the $j^{\mathrm{th}}$ frequency mode corresponding to the signal and the idler photon. $\omega_0$ is the degeneracy frequency of the SPDC process, $\Delta\omega$ is the FSR between modes and $\Pi$ is the spectral shape of each mode. We discretize the system in bins of 25 GHz bandwidth and the same FSR over the whole C-band, yielding the state: 
\begin{equation}
    |\psi_d\rangle = %\frac{1}{\sqrt{d}}
    \sum_{j=1}^{d} \alpha_j %e^{i\phi_{j}}
    |j\rangle_s |j\rangle_i
\end{equation}

The term $\alpha_j$ refers to the phase and amplitude of the mode and is determined by the spectral characteristics of the source. For a very broad and uniform spectrum as here, $\alpha_j\approx1$. Although after propagation, each photon pair corresponding to mode $j$ accumulate different phases due to material dispersion. 

In a quantum key distribution (QKD) scenario, the frequency-entangled photon pairs emerging from a single optical fiber can be distributed into different paths, e.g., via wavelength- or polarization demultiplexing. In our experiment, we use only one device simultaneously for both photons, but in principle, manipulation would be just as easily possible at two separate locations. 
To reveal the entanglement content, we use a commercial pulse shaper to control the phase and amplitude of each of the frequency modes \cite{kuesOnchipGenerationHighdimensional2017}, and electro-optic modulation to achieve mode superpositions \cite{olislagerImplementingTwophotonInterference2012} (see Experimental Section).\\

\subsection{The superposition basis}
\begin{figure}[t]
\centering
\includegraphics[width=0.47\textwidth]{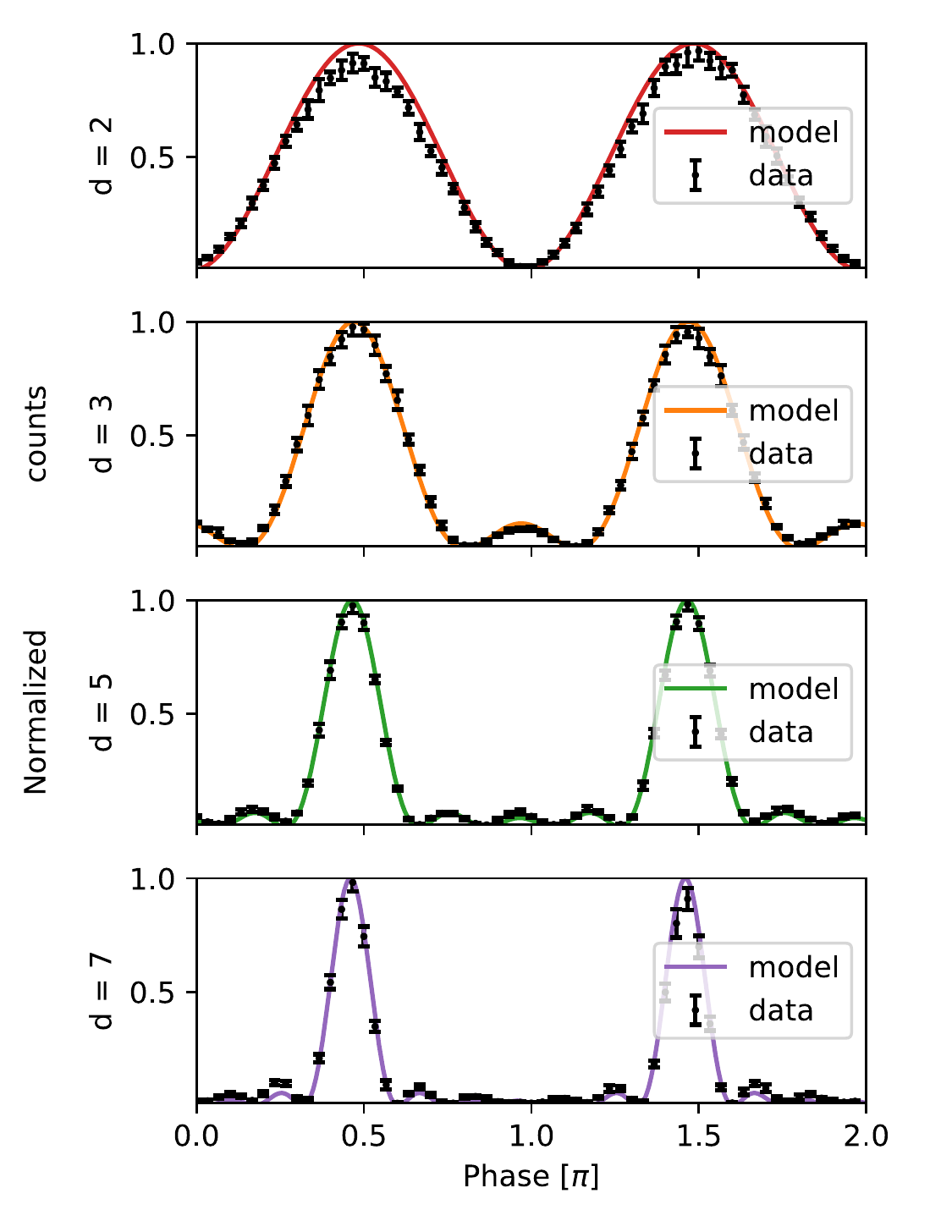}
\caption{Bell-type tests in the frequency domain for dimensions $d = 2, 3, 5, 7$. We chose the modes centered on $|6\rangle$ for the signal and the idler photons.
}
\label{fig:BellTest}
\end{figure}

For either quantum state characterization via full state tomography (FST), evaluation of Bell-type tests, or implementation of QKD protocols, measurements in superposition bases are fundamental to uncover quantum correlations and statistics of the state. 
The eigenvectors of these bases may be the superposition of some or all elements of the computational basis, here the frequency basis, with certain phases for each mode. 
Here, we show that with standard levels of RF signal amplification ($P_{\mathrm{max}}=26$ dBm), it is possible to perform a full superposition of up to 7 modes with a low contribution of accidental coincidence detection events. 
To prove this, we performed high-dimensional Bell-type tests, also known as CGLMP \cite{collinsBellInequalitiesArbitrarily2002}, based on the CHSH inequality for two-photon qubits \cite{clauserProposedExperimentTest1969}. Instead of measuring quantum correlations only for fixed phase settings, we performed a phase scan for all contributing modes \cite{thewBellTypeTestEnergyTime2004}. The measurement projector we use for each photon is:
\begin{equation}
    |\Psi_{proj}\rangle = \frac{1}{\sqrt{d}} \sum_{j=1}^{d} \left(e^{ij\theta_{s,i}}|j\rangle_{s,i}\right)
\end{equation}
where $\theta_{s,i}$ is the phase applied to the signal or the idler photon, and we use $\theta_s = \theta_i$. The phase of interference depends on the sum of the signal and idler phases. By scanning their phase, we obtain the quantum interferences shown in Fig. \ref{fig:BellTest} for dimensions $d=2,3,5,7$. The visibilities are 96.7\%, 97.7\%, 98.1\% and 98.2\%, respectively, without fitting or subtraction of accidental coincidences and much above the thresholds 70.7\%, 77.5\%, 84.6\% and 88.3\% to rule out hidden variable theories %assuming white noise
\cite{thewBellTypeTestEnergyTime2004}. 
To perform these measurements, we have selected states centered on the $6^{\mathrm{th}}$ mode. 
Electro-optic modulation shifted photons from the neighboring modes into the $6^{\mathrm{th}}$ mode, and demultiplexing filters (DEMUX) postselected the superposition state. 
Our constrained RF signal only allowed a limited efficiency of the frequency modulation of photons; thus, added mode loss by the pulse shaper provided an equal contribution. Even dimensionality can also be evaluated by using the same parameters as here. 

\subsection{Higher dimensionality certification}
While these CGLMP tests provide a (partially) device-independent certification of entanglement and demonstrate the reliability of our devices, they do not easily test for actual entanglement dimensionality, i.e. the dimension of entanglement needed to reproduce the correlations.
Generically, to demonstrate even higher levels of entanglement, one would need to project distant spectral modes into superposition states, which may be limited or physically impossible to perform. 
In our large frequency space and current setup, this would imply unreachable RF power levels for electro-optic modulation.
Although similar practical limitations are to be expected for any type of encoding, there exist other modulation methods that can provide frequency shifts of several nanometers \cite{joshiFrequencyMultiplexingQuasideterministic2018, zhuSpectralControlNonclassical2022}.
On the other hand, common entanglement certification techniques, such as FST, are expensive procedures that require measurements in at least $(d+1)^2$ bases for bipartite systems, an equivalent of $d^2(d+1)^2$ single-outcome measurement settings. The obvious consequence is that the number of total measurements increases rapidly with dimension. 
Adapting the techniques developed for the time-bin domain, we show that it is sufficient to characterize the quantum coherence of a few 2D subspaces to certify a high dimensionality of entanglement in the frequency domain.

\begin{figure}[ht]
\includegraphics[width=0.49\textwidth]{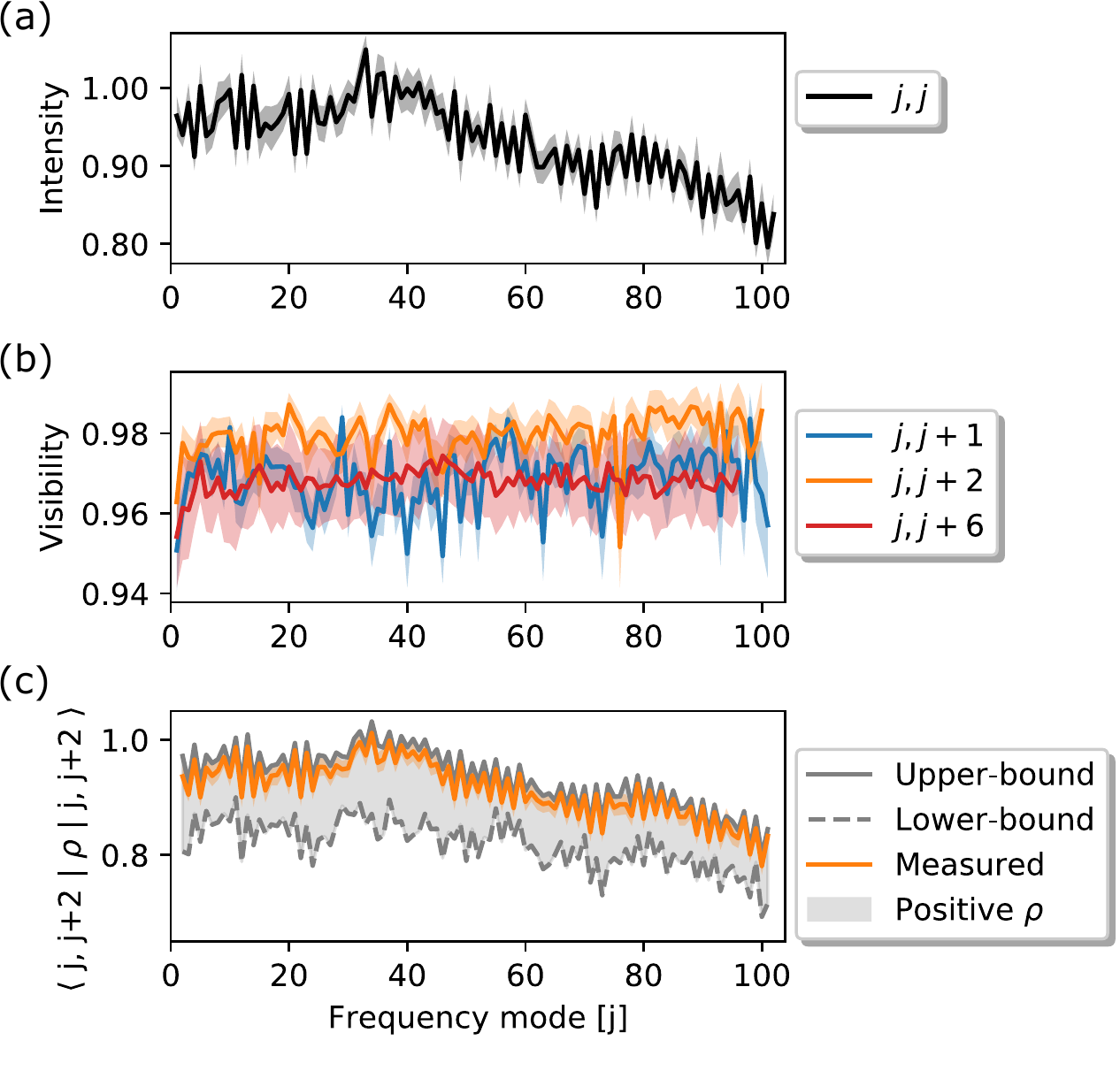}
\caption{(a) Intensity probabilities and (b) visibilities of the two-photon interferences for the 2D subspaces of modes $|j,j+1\rangle$, $|j,j+2\rangle$ and $|j,j+6\rangle$. (c) Comparison of measured data with the expected lower bound with our method. Note that, from all possible values of elements that produce a positive density matrix $\rho$, we take the worst scenario.}
\label{fig:Performance}
\end{figure}

The certification process is structured into two main blocks: the measurement of some elements of the density matrix $\rho$ and the posterior lower bound of the remaining unknowns. Measurements on the computational basis, that is, the frequency basis, can be performed with standard filters and are, in fact, directly related to the diagonal elements of the density matrix $\langle j, k| \rho |j, k\rangle$.
Only this characterization step would usually require $d^2 = 10404$ filter settings, which can take arbitrary long times. 
Here, we propose making good use of the frequency parallelization and bucket detection of all uncorrelated frequencies with a single and broadband filter setting $\sum_{k\ne j} \langle j, k| \rho |j,k\rangle$ (i.e. 101 modes $\times$ 25 GHz), while still measuring the correlated set $|j,j\rangle$ with narrowband filter settings (25 GHz). 
This method reduces the high number of measurements required to only $2d$. 
The results of the maximum frequency correlation are shown in Fig. \ref{fig:Performance}a. 
The background noise measured for uncorrelated frequencies originates from accidental coincidence detection events due to the high number of single counts and imperfect filters. 
The detected coincidence to accidental ratio (CAR) averaged throughout the space amounts to $1.4\times10^3$. 

Outside of the diagonal of $\rho$ we find two types of elements: those close to zero due to energy conservation, as observed from the computational basis measurements and upper limited by the accidentals in our system, and those non-zero elements $\langle j,j | \rho | k, k \rangle$ that indicate the coherence of the entangled state. 
We now measure some of the coherence elements to which we have access with our measurement system. 
They can be estimated with the mode amplitude from the computational basis and the strength of neighboring interference.
That is, we only need to measure the quantum interference between any two modes of the whole system. 
We measure the interferences for the first, second and sixth neighboring modes corresponding to the terms $\langle j,j|\rho|j+i,j+i\rangle$ for $i=1,2,6$. 
The dispersion shifts the interference patterns proportional to the frequency distance with respect to the center wavelength \cite{imanyCharacterizationCoherentQuantum2018}. 
Knowledge of the exact dispersion values would allow us to directly measure the maxima and minima of the interference patterns. 
For an unknown exact value as in our case, we perform a dispersion calibration, a full phase scan of the interference for different frequency modes. 
%We measure 40 points for modes $j=1,21,41,61,81$ and their respective neighbours. 
Finally, to collect statistics, we record 60 samples of the maxima and minima of the expected interference for each subspace and calculate the visibility. 
The total number of filter settings (not considering the dispersion calibration) are now $2(d-1) + 2(d-2) + 2(d-6) = 6(d-3)$ and the results of all recorded visibilities are presented in Fig. \ref{fig:Performance}b. 
The average visibility for the 102 modes with 1st, 2nd, and 6th neighbors are, respectively, 96.85(7)\%, 97.94(5)\% and 96.8(1)\%.

\begin{figure}[ht]
\centering
\includegraphics[%height=0.62
width=0.45\textwidth]{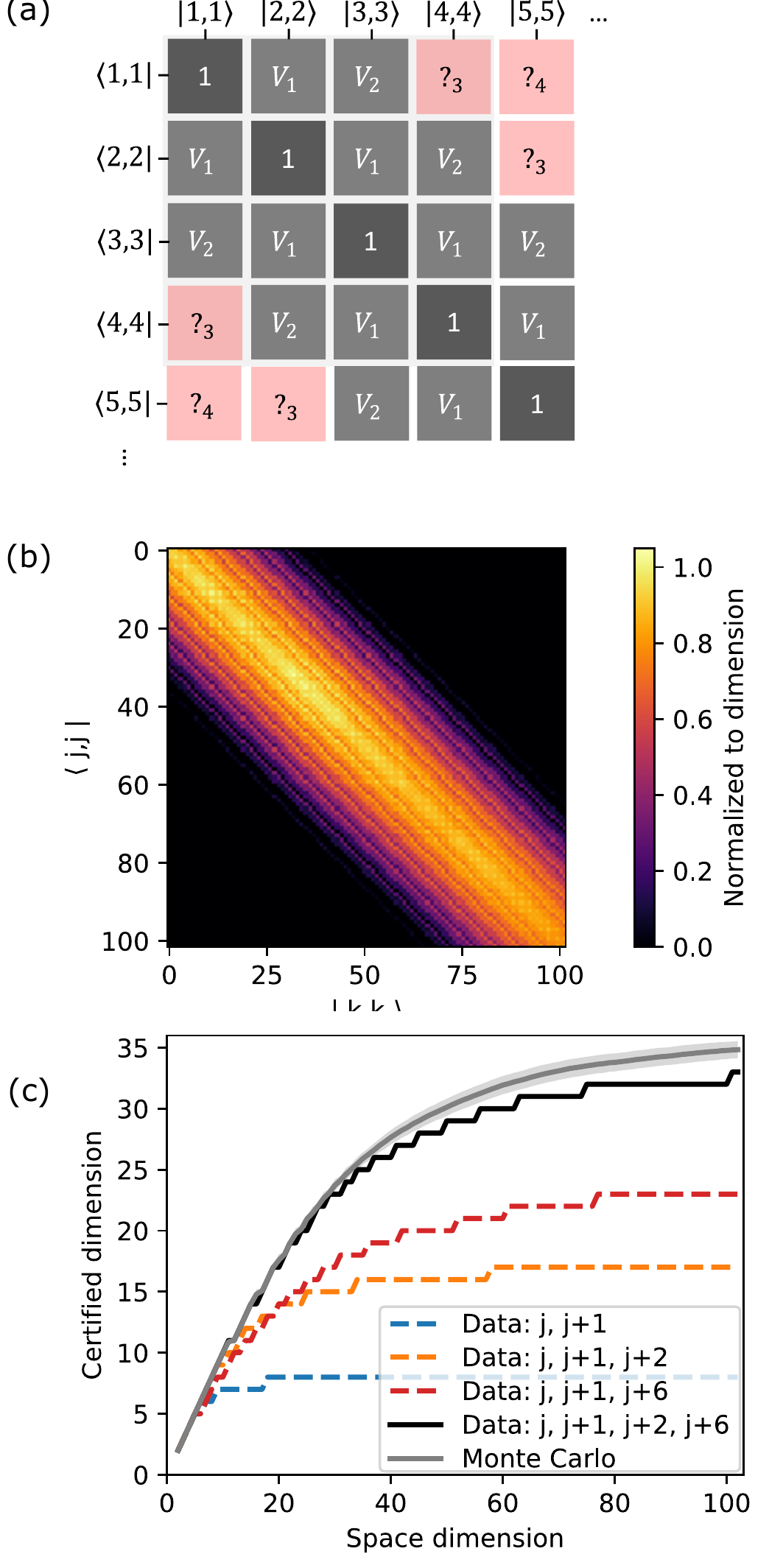}
    \caption{
(a) Simplified representation of the density matrix, considering only the elements $\langle j,j|\rho|k,k\rangle$. While some visibilities $V_{j}$ can be measured, the remaining unknowns in red are lower-bounded with our method.
(b)
 Lower bound of the density submatrix $\langle j,j | \rho | k,k \rangle $ and (c) minimum certified dimensionality of our system, according to the measured data. Notice that for low space dimension up to $d = 11$, we can certify the maximum dimensionality with very few measurement settings.}
    \label{fig:results}
\end{figure}

The fact that these high-contrast interference is preserved over the whole investigated space indicates that the quantum state at hand is very close to a maximally entangled state. However, in potentially adversarial scenarios, such as QKD, we do not want to make any assumption on the distributed state. Thus, we proceed with a rigorous analysis to demonstrate entanglement. Indeed, these strong coherences allow us to finally certify a large amount of entanglement without further measurements. 
Similarly to the methods proposed for entanglement certification in the time domain \cite{tiranovQuantificationMultidimensionalEntanglement2017,martinQuantifyingPhotonicHighDimensional2017}, we lower bound the remaining unknown elements $\langle j,j | \rho | j+i, j+i \rangle$ (see Fig. \ref{fig:results}a) by using the fact that the density matrix must be positive semidefinite to represent a valid quantum state. 
Thus, every element, submatrix, and therefore subdeterminant of $\rho$ must be positive or equal to zero at the very least. 
The magnitude of unknown density matrix elements can thus be lower-bounded iteratively by solving $3\times3$ subdeterminants, composed of measured or estimated parameters and one unknown, and we keep the largest bound extracted from all combinations of submatrices. 
The resulting submatrix $\langle j,j| \rho | k,k \rangle$ is shown in Fig. \ref{fig:results}b.
Notice that this method causes a rather fast loss of information, many of the elements are still lower-bounded to zero, yet it is sufficient to certify high-dimensional entanglement.
To visualize this loss, we display in Fig. \ref{fig:Performance}c the measured quantities for $\langle j,j|\rho|j+2, j+2\rangle$ and the calculated bound if we would not use those measurements.

We can now compare the lower-bounded density matrix with a target state $|\Phi\rangle$ by computing the fidelity $F(\rho,\Phi) = \mathrm{Tr}\left( \sqrt{\sqrt{\Phi}\rho\sqrt{\Phi}} \right)^2$. 
There exists an upper bound for the fidelity of any state of Schmidt rank $k\leq d$ \cite{ficklerInterfacePathOrbital2014,bavarescoMeasurementsTwoBases2018}. 
Fidelities with the maximally entangled state above the threshold $B_k(\Phi)=k/d$ indicate a dimensionality of at least $k+1$.
We thus choose a maximally entangled state as the target state $|\Phi\rangle=1/\sqrt{d} \sum_{j=1}^{d}|j,j\rangle$, iteratively calculate the fidelity for $d = 2$ to $d = 102$ and compare it with the threshold values for varying Schmidt rank. 
The final certification is plotted in Fig. \ref{fig:results}c, where we also show the dependence on the amount of measurement data. Certainly, collecting the coherence information from more distant modes helps to improve the certification. By using all the measured data, we find at least 33 entangled modes in a space of 101 to 102. 
To estimate the error of the certification, we perform Monte Carlo simulations with our data error. On average, the statistical method yields an even better certification, with 34.8 +- 0.7 entangled modes in a space of 102.
It is worth emphasizing that with only these very few measurement settings, $2d$ on the computational basis and $\sim 6d$ on 2D subspaces, we are able to certify 11 dimensions in a space of 11 modes. Higher visibility would directly increase the amount of certified entanglement per number of modes, and further measurements that could fill more elements of the density matrix would also improve the certification. 
\vspace{0.6cm}

\section{Conclusion}
The time-frequency domain straightforwardly provides entanglement of very high dimensionality simply owing to energy conservation in spontaneous parametric nonlinear processes. However, the characterization of this entanglement is a key challenge. To date, high-dimensional frequency entanglement has never been completely certified beyond a few dimensions. In our work, we have shown methods on how to characterize these quantum states that unquestionably possess a huge dimensionality. Similarly as in the time domain, the limits on the dimensionality for continuously pumped processes depend on the resolution of our devices, here the spectral filters. 

In this work, we have shown full superpositions up to $d=7$ and great interference visibilities over the full scanned spectrum. The used subspaces are easily exploitable with few fiber-integrated optical components and can be implemented nonlocally in quantum communication tasks. Due to physical measurement constrains, limited amount of information is accessible from these large states. Nevertheless, we have finally certified the full 11-dimensional entanglement in a subspace of 11 modes with only $\sim 8d$ measurement settings, and at least 33-dimensional entanglement in our frequency space of 102 frequency modes.

These results hold great promise for further enhancing the information capacity of quantum information processing. Entanglement in the frequency domain can also be used in combination with entanglement in time, path and polarization DOFs, leading to huge state spaces. The main building blocks and techniques can be readily combined with established spatial and temporal mode approaches and thus provide the near-term prospect of more efficient hybrid time-space photonic information processors. 

We hope that these results on the quantification of quantum coherence in a massive-dimensional frequency space will motivate the community to use the frequency DOF beyond wavelength multiplexing schemes. Moreover, the further development of photonic technology, in particular low-loss electro-optic modulation and wave-shaping technologies, could render frequency encoding a viable contender for near-term photonic quantum information processing with massive bandwidth.

\section{Experimental section}\label{Section:Exp}
To generate the frequency-entangled state, a commercial second harmonic generation (SHG) module was used, a 40 mm type-0 ppLN waveguide (Covesion). 
To pump the nonlinear process, a standard continuous-wave telecom laser was upconverted with a second SHG module to better align the SPDC wavelength to ITU channels. 
To process the frequency entanglement, a telecom pulse shaper (Waveshaper 16000A) was used to select specific frequency subspaces for the signal and the idler photons and to tune their relative phases. This device limits the operational frequency range of the source to 40 nm of the telecom C-band only. 
This space was then divided into 102 frequency modes for the signal and the idler photon, with a standard free spectral range (FSR) of 25 GHz and the same bandwidth. 
This leads to a state space of $102\times102$ dimensions. 
Note that subspace selection was an actual frequency discretization procedure, and it could be employed as a resource for flexible bandwidth allocation in reconfigurable QKD networks. 
The photon frequencies were then modulated with an electro-optic modulator, driven by a radio-frequency sine with the same frequency as the FSR. 
This sideband modulation technique allowed us to distribute photons into neighboring modes according to Bessel function amplitudes \cite{capmanyQuantumModelElectrooptical2010}. 
Choosing a low FSR and bandwidth allowed slower modulation signals than in earlier work, where FSR values of 200 GHz \cite{kuesOnchipGenerationHighdimensional2017} down to 50 GHz \cite{imany50GHzspacedCombHighdimensional2018} were used. Lower FSR would be better suited to increase dimensionality, but would eventually be limited by the resolution of the available optical filters and the number of photons per fraction of spectral bandwidth. Lastly, due to photon scattering into distant spectral modes, postselection was necessary to measure the right superposition states. To reduce the noise that imperfect filters may introduce, a tunable and narrow DEMUX of 22 GHz bandwidth centered at the corresponding signal and idler frequencies was used prior to coincidence measurement. The colorful illustrations at the bottom of Figure \ref{fig:Setup} depict the sideband modulation and photon scattering.

\medskip
\section*{Funding} \par 
This work was supported by the Fraunhofer Internal Programs under Grant no. Attract 066-604178, by the joint program of the Natural Sciences and Engineering Research Council of Canada (NSERC) and the European Comission under Grant no. HyperSpace 101070168, and by the Federal Ministry of Education and Research of Germany (BMBF) through the project Quantum Photonics Labs (QPL). The presented results were partially acquired using facilities and devices funded by the Free State of Thuringia within the application center Quantum Engineering. M.C.P is part of the Max Planck School of Photonics supported by BMBF, Max Planck Society, and Fraunhofer Society.

% References
\medskip

% Use the following code if you wish to generate your bibliography with BibTeX;
% replace the string "MSP-template" below with the name(s) of
% the BibTeX data base(s) you want to use.
% The resulting bibliography-output (the content of the .bbl file)
% must be pasted back into this file before submission.
% Please also include your BibTeX data base file(s) in your submission
% so that we can re-run BibTeX if necessary.
\bibliographystyle{MSP}
\bibliography{MyBiblio_Rev}

\end{document}